# Cyberwar Strategy and Tactics
## An Analysis of Cyber Goals, Strategies, Tactics, and Techniques

Laura S. Tinnel, O. Sami Saydjari, and Dave Farrell

*Abstract – Cyberwar strategy and tactics today are primitive and ad-hoc, resulting in an ineffective and reactive cyber fighting force. A Cyberwar Playbook is an encoding of knowledge on how to effectively handle a variety of cyberwar situations. It takes a troubleshooting approach and defines the cyber tactics, techniques and procedures one may employ to counter or avert cyber-based attacks. It provides focus and clarity in time of chaos allowing a clear path of response to be chosen*.

## I.  Introduction

Millennia of warfare experience have yielded effective strategy and tactics in conventional or so-called kinetic warfare.  Without strategy and tactics, soldiers are an ineffective fighting force, milling about aimlessly, unsure of their overall goals, their immediate objectives, how to cooperate amongst themselves, and what to do when facing an enemy.  Soldiers are relegated to reacting to the situation at the moment, either seeking out the enemy randomly or reacting to an attack with each individual deciding what to do when he happens to perceive the attack.

Such chaos is the current situation in today's networked computer systems, which we will refer to as cyberspace. As we evolve from no defenses, through static defenses, toward the emerging highly configurable mechanisms of next generation systems [1], we need an understanding of how to effectively configure and orchestrate these flexible mechanisms.

Without such strategy and tactics, we have no clear idea about what configurations are effective against the variety of attacks that exist, and more importantly, against those that are continuing to evolve.  This becomes especially important as attackers move from attacks involving just a few steps to well-thought-out strategic campaigns involving many simultaneous battles toward some specific goal.
Adversaries can and will plan and execute strategic cyber attack campaigns against the United States.  The adversary is at a strategic advantage, having arbitrarily long periods of time to formulate plans and the luxury of choosing a beneficial execution time that includes the element of surprise.

Defenders, on the other hand, are hard pressed to recognize anything more than the lowest level campaign step and cannot see the bigger picture. At best, they are relegated to tactical reactions to adversary stimuli on an event-by-event basis.  At worst, their reactions are not even tactical, but are knee-jerk and potentially harmful.  It is impossible to get ahead of the adversary when taking such a haphazard, reactionary defense posture.

Defenders need to actively capture their experience base with real attacks to advance to strategic defense.  They need to use their time wisely to begin working out anticipatory defense strategies for likely cyber attack scenarios.

## II.  Definition

A *Cyberwar Playbook* is an encoding of knowledge on how to effectively handle a variety of cyberwar situations. It uses concepts such as deception, confusion, stimulation, and blockading and defines the cyber tactics, techniques and procedures one may employ given various strategic goals.  It is independent of any specific network infrastructure and is not intended for use in building a static defense plan. Rather, its intended purpose is to aid human defenders during *active* cyber attack situations. It is used to quickly determine the best actions to take when faced with a given situation. Thus it contains the adversarial moves one might expect to see and the countermoves believed to be effective against those moves.

We expect the playbook to contain a list of plays that are arranged by situation or strategies and used by planners to achieve goals for a commander. The playbook will hold templated parameters that will assist users in selecting the optimal play for the given situation.

We expect the playbook to be a living document, updated constantly to capture our continuously evolving knowledge of attacks and defenses. We also expect that, because of differences in responsibility, perspective, and authority, there will ultimately be multiple playbooks. For example, we expect different echelon levels to need unique sets of plays, as will parallel organizations.

## III. PLAYBOOK APPROACH

### A. Understanding The Available Moves

To create a cyberwar playbook, we must first understand the stratagem building blocks or *possible moves* that are available. It is important to note however that these stratagem building blocks in and of themselves are not strategic. Instead, it is the reasoned application of one or more stratagems in accomplishing higher-level goals that is strategic in nature. We thus need to understand the situations in which the stratagems should be applied and how. We can begin to predict and choose the most effective stratagem for a given situation as we become more experienced. Example stratagems include:

```
Fortify     Dodge
Deceive     Block
Stimulate   Skirt
Condition   Monitor
```

Stratagems may also have sub-stratagems. Examples are:

```
Deceive.Chaff       Block.Barricade
Deceive.Fakeout     Block.Cutoff
Deceive.Conceal     Monitor.Eavesdrop
Deceive.Feint       Monitor.Watch
Deceive.Misinform   Monitor.Follow
```

These stratagems are very high level and can be supported through many tactical means. Each building block defines a stratagem and contains one or more possible tactical implementations for that stratagem, including requirements, goals that may be satisfied using the stratagem, caveats, example uses, and possible countermeasures. Table 1 shows the elements of a building block.

**Table 1: Example Stratagem Building Block**

| Stratagem | Dodge |
|---|---|
| Description | Make sudden movement in new direction; move to and fro usually in irregular and unpredictable pattern |
| Example Tactical Implementation | Change IP address of target host so attack packets do not reach host; de-list in DNS and use local host file for resolution. |
| Infrastructure Properties Where Useful | Only small number of hosts / users need access to target host. |
| Technological Requirement | Mechanism to securely push or update hosts file across the net. |
| Goals Which May be Satisfied | Maintain reliable service of critical host. |
| Example Attack / Adversary Properties Where Helpful | Network based attacks from outside the LAN or where adversary has no means to access the updated hosts files. |
| Effects on Adversary | Adversary packets no longer reach host. Disrupts attack until adversary discovers new address. |
| Limitations and Assumptions | Must detect IP of target and take specific action. May only work for short period of time. |
| Implications | Users of services will be cut off from service if host files not distributed or internal DNS not updated. |
| Example Red Use | Change IP of host that is target of IA control from command center. |
| Example Blue Countermeasure | Tripwire firewalls and switches carefully; be able to quickly change firewalls and switches; have IDSs look for unexpected IP translation. |

The full set of building blocks is captured in a separate document [2]. The remainder of this paper focuses on the issues of how and when to apply these stratagems in cyberwar situations.

### B. Play Development Approach

We determined that in order to consider a strategic cyberwar attack, we needed to postulate a concrete mission upon which the defender was focused and working toward. We found that without doing so, it was difficult to maintain our thoughts at the strategic level and that it was almost natural to begin applying low-level reactionary responses to system level events as opposed to tactics. This is problematic because the application of tactics is important to achieving strategic goals. Further, Cyberwar offensive strategies have to do with attacking defender missions; attacking defender systems is simply a means to that end.

We thus postulated a military troop deployment mission[†]. In looking back over our analysis, it seems that there is nothing specific to our particular choice of mission. This is good because in order for a playbook to be widely useful, it must have application to a broad set of missions. Still, we believe it beneficial to have a concrete mission in mind as one thinks through the strategy.

We took a simple analytical approach and first postulated some reasonable adversary (Red) counter-missions to the defender's (Blue) mission. Of those we chose one. There are certainly many alternatives that can and should be considered to make the analysis complete. For the Red mission, we examined the stratagem list and considered moves we believed to be effective in accomplishing the adversary's counter-mission. We then applied those stratagems to our particular situation. For each stratagem, we considered how a defender might detect which stratagem was being used and how he might defend against it. We then examined the Red side again. By engaging in only a few iterations of move/counter-move we could often determine when one stratagem might be more effective than another or identify moves that could result in recursive or other undesirable situations. This information is useful in pruning the space of viable stratagems and guides the selection of the most optimal choice for each attack situation.

---

[†] Done loosely using the DARPA Cyber Panel program's Grand Challenge Problem for background context. Because the problem description is unpublished, we provide sufficient detail to give the reader the general context.

In what follows, we present our rudimentary analysis progression, admittedly far from complete. We then move to a section on insights and conclusions. We found that our analysis stimulated insights into some possible defense strategies that might not have been obvious otherwise, and that could prove effective even without significant advancement of the underlying technology.

IV.  RANGE OF CYBER ASSETS

Before going on to discuss the application of stratagems, we take a moment to describe the substrate on which cyber actions take place - the cyber assets. Cyber assets include different entity categories that heretofore may have not been explicitly considered in a strategic cyber defense context. The cyber assets may be categorized as follows: (1) *Intelligence Gathering Assets*, (2) *Effect On Resources*, (3) *Cyber Assets Defense Posture*, and (4) *Cyber Asset Status*.

*Intelligence Gathering Assets* refers to the collection of systems that are to gather intelligence on the adversary. Put simply, this is a matter of watching the watchers to see what they are looking at, or even that they are more intensely looking at something. To use a physical analogy, one can look in the parking lot of an adversary's intelligence agency and see that it is full on Sunday and conclude that something significant is happening. In the cyber world, this is a matter of tracking the activities of known cyber intelligence programs and servers on the system that receives cyber event reports from the field.

*Effect On Resources* refers to increased activity not just of the intelligence gathering assets of the adversary, but of the effects on your own system of those intelligence gathering activities. For example, one might expect network mapping and probe rates to increase before an attack happens.

*Cyber Assets Defense Posture* refers to looking at the defense posture of an adversary, including his INFOCON level, the tightness of his firewall and authentication policies, and such. Often, we might expect that an adversary will "batten down the hatches" before he throws the first punch in a battle.

*Cyber Asset Status* refers to potential realignments and positioning of functions. For example, commercial sites might suddenly go offline and be used for intelligence purposes. High-priority functions may be re-positioned to higher security enclaves. Certain open services may be discontinued.

V.  APPLICATION OF STRATAGEMS

As stated above, the Red mission is to interfere with the ability to deploy troops, specifically to cause a two-week delay from any arbitrary starting time.

Application of the stratagem building blocks is considered a strategic activity to support the overall plan to reach a desired goal. At a level below that, there are tactical steps required to achieve the chosen stratagem. For example, a stratagem to disrupt a target system requires an attack tree that considers the specific function and topology of the target system required to execute the stratagem.

Intelligence gathering is one of the first stratagems to be applied – one not traditionally considered in the cyberwar arena. Both sides are constantly monitoring one another and both sides know it. Intelligence gathering is performed at two different levels. There is standing or generic intelligence gathering on activities that we expect might be useful to us in the future. Such intelligence gathering is a generic strategic move against a generic situation. Then there is focused intelligence gathering that can be done with respect to a specific situation. Often such focused gathering counts on infrastructure that has been pre-placed for generic gathering purposes or in anticipation of a focused gathering requirement.

Standing intelligence gathering is always being performed and can be used in preparation for something else; for example, the invasion of country X, where you obtain some intelligence on that country as a regular part of intelligence gathering duties. You can only hope that your standing intelligence gathering activity is collecting information useful for such a specific situation.

Focused intelligence gathering is often one of the first steps of an adversary as a precursor to future events. This may include out-of-band techniques (e.g. HUMINT) that cannot be detected within the system. However, intelligence gathering using cyber assets is at least potentially detectable within the target cyberspace itself.

This activity includes the stratagem block `Monitor.Watch`, as well as tactical activities to support it, such as gathering data from the deployed sensors, reconfiguring current sensors and activating new sensors.

One may assume some level of deception as the adversary may purposely expose false information. As an example, Red may retrieve massive amounts of benign data in an attempt to mask the real data that they are gathering from Blue's cyber assets.

They may then apply a prioritization scheme for next steps based on the perceived effectiveness of this deception.

The cyber assets used by Red to perform this function may also change posture during this activity, for example applying the `Stealth` stratagem to minimize exposure or using other misdirection techniques. Red may also apply the `Fortify` stratagem to their own assets just in case Blue decides to actively penetrate Red's system to learn the actual target of Red's focused intelligence activity.

As an initial cut, for a given scenario, we propose each stratagem take a range of 3 intensity levels: low, medium, high.

### A. Opening Salvo: Red

Red uses the `Monitor.Watch` stratagem to gather intelligence to understand how troops are deployed. The initial knowledge state is fed by the standing order to gather intelligence and is presumed to be non-zero.

As Red's `Monitor.Watch` activity increases in level, Blue's `Monitor.Watch` activity should detect this and correspondingly increase its activity level in response.

Red then uses a `Deceive` stratagem to fool Blue `Monitor.Watch` activities. Red deceives by gathering focused intelligence on multiple theatres of military operation.

Red also deploys decoy `Monitor.Watch` assets, thus increasing the number of apparent intelligence gathering assets. An example decoy would be Red programs that randomly probe Blue systems and perhaps also send encrypted nonsense traffic back to intelligence headquarters computers.

Red also applies `Fortify` to the actual intelligence gathering assets in anticipation of Blue beginning to attack Red's `Monitor.Watch` system to learn its target.

Note that Red's fortification activity may take a long time to accomplish; for example, an increase in Defense Condition (DEFCON) may take days due to many manual processes.

Red then also applies `Fortify` to other non-intelligence gathering cyber assets as in preparation for cyber war.

### B. Blue's Response Initiatives

Blue's `Monitor.Watch` detects an increase in Red's `Monitor.Watch` and `Fortify` activities.

Blue initiates a focused intelligence gathering activity to learn the nature of Red's `Monitor.Watch` actions. To do so, Blue starts two actions. First, Blue attempts to infiltrate Red's cyber intelligence system to directly learn its targets by examining captured programs. Second, Blue creates a `Fish Bowl` to monitor Red's intelligence gathering activity to gain tactical information to infer Red's strategy. Both moves are uncertain in their probability of success and in the time they will take to execute.

Blue then moves its worldwide `Fortification` level from low (presumed peacetime level) to medium in anticipation of a possible attack from Red based solely on the increased activity seen so far.

Blue also uses a `Deceive.HoneyPot` stratagem. Blue deploys multiple fake systems covering multiple mission spaces (e.g. logistics, command and control, intelligence, etc.) to gain clues as to Red's real target. This move does double-duty in that it also employs the `Deceive.Decoy` stratagem by deflecting Red from Blue's real cyber assets. The move also buys time to allow Blue `Fortification` to take place.

The analysis then continues in the same manner with Red's countermoves and so on. Several layers of analysis should be conducted to facilitate planning and avoid blind-alley moves that either lead nowhere or hamper Red or Blue with respect to their goals. As weighting and pruning strategies become viable in the information assurance domain, they should be applied to the possible moves.

### C. Execution Phase - Weaponry

Once the planning analysis is done, as described above, both Red and Blue need to create specialized programs and tools to execute the specific steps prescribed by the planning process (and react to unforeseen actions during execution). The programs and tools are custom in that they must be tailored to the target system environment and the particular goal that the programs are trying to accomplish. Unlike soldiers, malicious code must be given very explicit instructions for the actions it is to take.

In our scenario, through cyber intelligence gathering and analysis, Red now knows how the system is used to deploy troops. Red also has likely discerned some of Blue's stratagems as Red observed Blue's responses to Red's moves. Red would likely next create cyber weapons (programs) for integrity and Denial of Service (DoS) attacks on Blue's systems to accomplish Red's delayed deployment goal.

Meanwhile, execution for Blue is more problematic. Blue's focused infiltration attempt is likely still ongoing and yet to be successful. It will be difficult for Blue to detect the start of Red's attack planning phase unless Blue can somehow expose Red's tools and techniques for testing their attacks before Red executes them.

Because of these challenges, Blue decides to insert a human insider into Red's organization. Although the move is immediately initiated, it may not yield results for a long time. Blue takes a gamble that the move will either have surprisingly early success or that the conflict will last long enough for insider placement to be worthwhile.

Once the detailed attack plans are complete and the cyber weaponry is built and field-tested, the attack execution phase begins. Red exploits cheap, openly available attacks first so that Blue might mistakenly attribute the attacks to one of Blue's many low-level adversaries.

Red would hold back more complex sophisticated attacks to hide their real capabilities and to avoid having their tools, tactics, and procedures matched against their "signature" profile early on.

### D. General Strategic Activity Sequence

We discovered as part of our analysis that we were following a general sequence of activity flow. The sequence is more tree-like and not strictly linear.

Figure 1 shows the notional timeline for the general sequence of activities we engaged in when applying strategy to cyber situations. Results from the standing cyber intelligence gathering automatically spawned focused cyber intelligence gathering and generic defense posture tightening. Results from focused cyber intelligence gathering automatically spawned focused defense posture tightening, asset repositioning, counter-attack planning, and weaponry countermeasures. Counter-attack planning spawned counter-attack execution. The cycle then repeated itself.

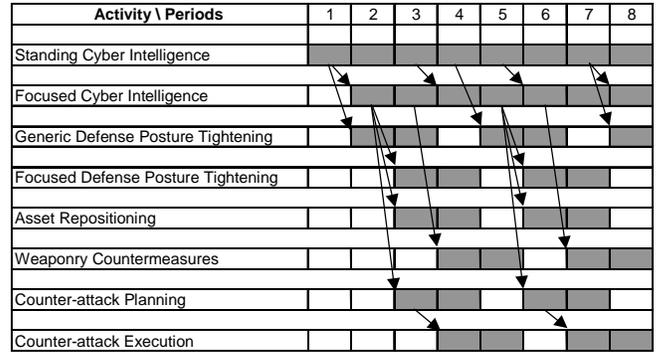

Figure 1: Timeline for Strategic Sets of Actions

We believe this sequence of actions would occur during any situation and scenario. Further analysis and experimentation is warranted in this area.

## VI. RELATED WORK

Zel Technologies has applied traditional Intelligence Preparation of the Battlespace (IPB) to the information domain. IPIB provides actionable, predictive information about probable adversary courses of action (COAs) prior to actual hostilities, and is generally viewed as a system defense design time utility. IPIB was shown through a white board experiment [3] to aid defenders in accurately predicting cyber attack targets based on mission utilization. This mission-based approach is similar to our first approach to playbook development and does assume that the enemy's goal is to always counter the defender's mission, which may not be the case. The IPIB process did not prove particularly useful in projecting enemy COAs that were strictly cyber based, but did help in predicting COAs involving combined kinetic and cyber activities, which are highly likely in a true war situation. We believe our and the IPIB efforts can be mutually beneficial. IPIB focuses strictly on prevention and detection. Missing from the IPIB work are the strategies and tactics that focus on deception and other concepts that are needed in "heat of the battle" situations. A cyberwar playbook can provide this needed input. IPIB could potentially be used to expose situations for which plays should be developed and encoded in the playbook.

Orincon Corporation is working on the application of game theory to defensive information warfare [4]. Early results on automated tree exploration and pruned searching are encouraging. Their findings on topics such as timing issues are consistent with this work. Fundamental work on this project serves to feed a more automated codification in the Orincon project. Early results on this project have been freely shared to encourage this synergy.

Research in the area of anticipatory planning was conducted jointly by researchers at Texas A&M and the US Military Academy at West Point. Anticipatory planning seeks to define a new approach to military planning and execution for Information Operations, one that accounts for the "chaotic nature of warfare in which possibilities appear and disappear." The idea behind this approach is to create a plan with multiple branches that address as many of the adversary's likely and dangerous options as possible. Defender actions are maintained for as many enemy actions as possible. Branches in the plan represent transitions to a new state based on enemy actions. Anticipatory planning for a branch is completed well in advance so that reactive planning is not necessary once a branch occurs [5]. Continuous plan monitoring occurs and adjustments to the course of action are constantly being made. This planning approach is not specific to any particular domain and could be applied to cyber warfare.

## VII. INSIGHTS AND CONCLUSIONS

Time matters. Often, when we consider red team attack trees, time is washed out of the equation because it is difficult to have a red team operate over years (or even months) or even simulate the operation of the system and the red team over years. Yet, at the strategic level timing is everything.

Strategic action outcomes may take indefinite amounts of time and the results may come too late to be useful. Examples here include developing an insider or lifecycle attack AFTER a situation begins. Developing insiders can take years. When a defender takes such an action, he is taking the chance that the outcome may not happen in time. He does not know in advance how long the campaign will take. He must weigh the cost of the action against the probability of success and the probability of the timing of the outcome being useful in the campaign in which he is interested.

The generalization of specialized actions into standing activities was an interesting theme that arose from our analysis. For example, we should have a standing capability to watch the adversary's cyber defense posture as a change in that posture may indirectly indicate preparation for a surprise adversary action.

Deception is an important strategy that is underutilized and not considered enough in today's cyberwar landscape. There are many subtleties in how to deploy decoys and how to detect their deployment. This area warrants focused attention.

Finally, there will likely be plays, particularly in the offensive realm, that require higher levels of authorization to invoke. It is important to capture and provide information on who is authorized to execute which plays so that time to authorization may be taken into account when planning the next move. The lower the level of authorization needed to invoke a play, the faster one can respond in an active cyberwar situation. It is further important to provide plays requiring different levels of authorization so that some actions may be taken while waiting on authorization to proceed with other actions.

## VIII. FUTURE DIRECTIONS

Various scenarios should be examined to further explore application of and validate the playbook concept. Areas for validation include formats for human and computer use, the play development method, and the overall concept and usefulness of a cyberwar playbook.

A workshop was held that resulted in several suggested template formats, one of which attempted to create a Tactics, Techniques, and Procedures (TTP) field manual for cyber attack situations. Interactive group discussions utilizing red and blue teams to explore alternate scenarios will help flesh out the template details and identify strengths and weaknesses of the initial human-use format. Engaging the play development process with a small number of mock scenarios will provide simple validation of the method. Actual field-testing should be conducted to prove and refine the encoded plays and the playbook concept as a whole. To field test, the playbook should be used in conjunction with current techniques and compared to actual results, not just conjecture. Lessons learned should be folded into the work and the cycle repeated. We expect fairly rapid convergence towards a stable core that requires only minor modifications (e.g., incorporating local terminology) in order to provide broad applicability.

User buy-in is critical to ensure that the end product correctly addresses a real need and is readily accepted. Close collaboration between the research and end user communities will instill ownership on both sides and ensure buy-in. A start to this was seen at the workshop where both communities were represented and a joint consensus was achieved on the need for a playbook and on the concepts it must contain. Acceptance from top to bottom is imperative; thus input should be sought from those affected by the decisions made using the playbook, not just the decision makers.

The research community should leverage the body of knowledge gained by collaboration with the end user community in order to achieve significant technological advances. Work must be conducted to identify and explore promising cyber representations of the playbook. Today, fully automated response in a cyber environment is not realistic. However, we believe that by capturing and encoding authorization levels in addition to situational information, we will provide the knowledge foundation needed to support incremental insertion of semi-automated capabilities as advances are made in this area.

Finally, although the intent of the playbook is to support national security by aiding the cyberwar fighter, it has clear applicability throughout the public business sector, particularly in defending the critical infrastructure.

The groundwork has been laid and the level of end user participation and enthusiasm generated by our initial work has verified the need. The playbook must now be created and carried forth into actual use, perhaps through experimentation and exercises at first, to validate the utility and to gain invaluable feedback for improvement.